\documentclass[letter, colorlinks=true, linkcolor=blue, citecolor=blue, urlcolor=blue]{aa}  

\usepackage{graphicx}          
\usepackage{txfonts}
\usepackage{lipsum}
\usepackage{caption}
\usepackage{subcaption}         
                            
\usepackage{lscape}             
\usepackage{placeins}          

\usepackage{xcolor}
\usepackage{comment}
\usepackage{tabularx}
\usepackage{hyperref}
\usepackage{stfloats} 
\usepackage{afterpage}
\usepackage{footmisc}

\begin{document}

    \titlerunning{Early X-ray emission as an observational diagnostic of gamma-ray burst progenitors}
    \authorrunning{A. Ierardi et al.}

   \title{Early X-ray emission as an observational diagnostic \\of gamma-ray burst progenitors}

   \author{Annarita Ierardi\inst{1}\fnmsep\inst{2}\fnmsep\thanks{Corresponding author: \texttt{annarita.ierardi@gssi.it}},
           Alessio Mei\inst{3},
           Gor Oganesyan\inst{1}\fnmsep\inst{2}\fnmsep\inst{4},
           Stefano Ascenzi\inst{1}\fnmsep\inst{2}\fnmsep\inst{3},
           Samuele Ronchini\inst{1}\fnmsep\inst{2},\\
           Marica Branchesi\inst{1}\fnmsep\inst{2}\fnmsep\inst{4},
           Biswajit Banerjee\inst{1}\fnmsep\inst{2}\fnmsep\inst{4},
           and Maria Edvige Ravasio\inst{5}\fnmsep\inst{6}\fnmsep\inst{7}\\
        }

    \institute{
    \inst{1} Gran Sasso Science Institute (GSSI), Viale F. Crispi 7, L’Aquila (AQ), I-67100, Italy\\
    \inst{2} INFN - Laboratori Nazionali del Gran Sasso, L’Aquila (AQ), I-67100, Italy\\
    \inst{3} INAF - Osservatorio Astronomico di Brera, via E. Bianchi 46, I23807 Merate (LC), Italy \\
    \inst{4} INAF - Osservatorio Astronomico d’Abruzzo, Via M. Maggini snc, I-64100 Teramo, Italy \\
    \inst{5} Institute of Space Sciences (ICE, CSIC), Campus UAB, Carrer de Can Magrans s/n, Barcelona, E-08193, Spain\\
    \inst{6} Institut d’Estudis Espacials de Catalunya (IEEC), Edifici RDIT, Campus UPC, Castelldefels (Barcelona), E-08860, Spain\\
    \inst{7} Department of Astrophysics/IMAPP, Radboud University, Nijmegen,
6500 GL, The Netherlands}

   \date{Received \today}

  \abstract
  {Identifying the progenitors of gamma-ray bursts (GRBs) from the properties of their early high-energy emission detected by wide-field satellites is crucial for advancing our understanding of GRB physics and for multi-messenger astronomy. Current classification methods, which rely primarily on prompt emission properties, exhibit large scatter and are challenged by recently detected events. In this Letter, we propose a new method to classify GRB progenitors using their early X-ray emission, which is detectable by current wide-field X-ray monitors. We performed a systematic time-resolved spectral analysis of the X-ray steep decay phase for a sample of 23 collapsar-driven GRB candidates detected by the \textit{Neil Gehrels Swift Observatory}. We fitted the spectra over the $0.3 - 150 \, \rm keV$ energy range using both a physical synchrotron model and the empirical Band function, allowing us to constrain the spectral peak energy and the bolometric flux. We find a correlation between the rest-frame synchrotron cooling frequency, $\nu_{c,z}$, and the isotropic-equivalent luminosity, $L_{\rm iso}$, that is clearly separated from the $\nu_{c,z} - L_{\rm iso}$ relation followed by merger-driven GRB candidates. A similarly distinct separation is found in the correlation between the rest-frame peak energy, $E_{p,z}$, and $L_{\rm iso}$ obtained with the Band model. Finally, we provide a prescription for classifying fast X-ray transients detected by the \textit{Wide-field X-ray Telescope} aboard the \textit{Einstein Probe} mission, as its narrow energy range ($0.5 - 4 \, \rm keV$) limits spectral modelling to a simple power law. We show that the photon index and X-ray luminosity of the steep decay spectra clearly separate collapsar- and merger-driven candidate events, making them powerful diagnostics of GRB progenitors.}

   \maketitle

\nolinenumbers
%%%%%%%%%%%%%%%%%%%%%%%%%%%%%%%%%%%%%%%%%%%%%%%%%%%%%%%%%%%%%%
\section{Introduction}

Gamma-ray bursts (GRBs) are the most luminous electromagnetic transients observed in the Universe. Their radiation is characterized by a short-lived and highly variable prompt emission phase, typically detected in the MeV domain, followed by a long-lasting afterglow observable across the electromagnetic spectrum. GRBs are extragalactic sources and are powered by ultra-relativistic, collimated jets. Based on the bimodal distribution of the prompt emission duration \citep{1981Ap&SS..80....3M,1993ApJ...413L.101K}, they have been suggested to originate from two distinct progenitor channels: the merger of compact objects \citep[short GRBs, $T_{90}\leq 2\, \rm s$;][]{1989Natur.340..126E} and the core collapse of massive stars \citep[long GRBs, $T_{90}>2\, \rm s$;][]{1993ApJ...405..273W}.

However, a firm identification of the GRB progenitor requires additional observational evidence. In particular, the detection of associated thermal transients can provide direct constraints on the progenitor: a supernova (SN) supports a collapsar origin \citep{1998Natur.395..670G}, whereas a kilonova (KN) indicates a compact binary merger origin \citep{1998ApJ...507L..59L,2017ApJ...848L..12A}. Another complementary approach is to study the GRB host galaxy: collapsar progenitors are generally associated with star-forming environments \citep{2006Natur.441..463F}, while merger-driven GRBs are found both in old and young stellar populations and at larger offsets from their host galaxies \citep{2011NewAR..55....1B,2013ApJ...776...18F}. Alternatively, gravitational wave (GW) detection can provide direct confirmation of the merger scenario, as demonstrated by the joint detection of GW~170817 and GRB~170817A from a binary neutron star (BNS) merger \citep{2017ApJ...848L..13A,2017PhRvL.119p1101A,2017ApJ...848L..14G,2017ApJ...848L..15S,2017ApJ...848L..12A}.

Identifying the thermal transient associated with a GRB through optical follow-up observations is challenging, as it requires rapid and accurate localization of the source, which is often hampered by the poor angular resolution of MeV instruments. Furthermore, the limited detection horizons of optical telescopes for SNe and KNe restrict their identification to relatively nearby events \citep{2026arXiv260726133L}, in contrast to the cosmological distances at which GRBs are detected. Similarly, the sensitivity of the LIGO-Virgo-KAGRA detector network currently limits the detection of GWs from BNS mergers to distances of a few hundred Mpc.

In the absence of such additional observational evidence, the classification of GRB progenitors relies on MeV prompt emission properties such as duration, spectral hardness, spectral lag, minimum variability timescale, and energetics \citep[and references therein]{2009ApJ...703.1696Z}. However, these diagnostics exhibit significant scatter, and no single method provides a fully reliable classification. This limitation is further highlighted by both past and recently detected events that challenge the traditional duration-progenitor dichotomy. These include GRB~200826A, a short burst associated with a SN \citep{2020MNRAS.493.3379R}, as well as GRB~211211A and GRB~230703A, two long GRBs associated with KNe \citep{2022Natur.612..223R,2024Natur.626..737L}.

In this Letter, we explore the possibility of classifying GRB progenitors using their early X-ray emission ($t<10^3\, \rm s$). We are motivated by our previous work \citep{2026A&A...708A.190I}, in which we found that the early X-ray emission of merger-driven GRB candidates follows tight spectral-luminosity correlations, which extend prompt emission relations \citep{2004ApJ...609..935Y,2025A&A...693A.156M} to lower energies. Here, we investigate whether similar relations also exist for the early X-ray emission of candidate collapsar-driven GRBs and whether they are distinct from those observed for merger-driven GRB candidates. This approach is particularly promising given the current availability of wide-field X-ray surveys from missions such as \textit{Einstein Probe} \citep[EP;][]{2022hxga.book...86Y}. 
{Moreover, X-ray surveys can typically detect lower fluxes than MeV instruments, allowing transients to be followed over longer timescales.} Throughout this work, we assume a standard cosmology with $H_{0} = 67.7\ \mathrm{km\ s^{-1}\ Mpc^{-1}}$,  $\Omega_{m} = 0.3$, and  $\Omega_{\Lambda} = 0.7$.

%%%%%%%%%%%%%%%%%%%%%%%%%%%%%%%%%%%%%%%%%%%%%%%%%%%%%%%%%%%%%%

\section{Sample selection and spectral analysis}
\label{sec:methods}

\begin{figure*}[t!]
   \centering
   % First subfigure: light curve
      \includegraphics[width=1\linewidth]{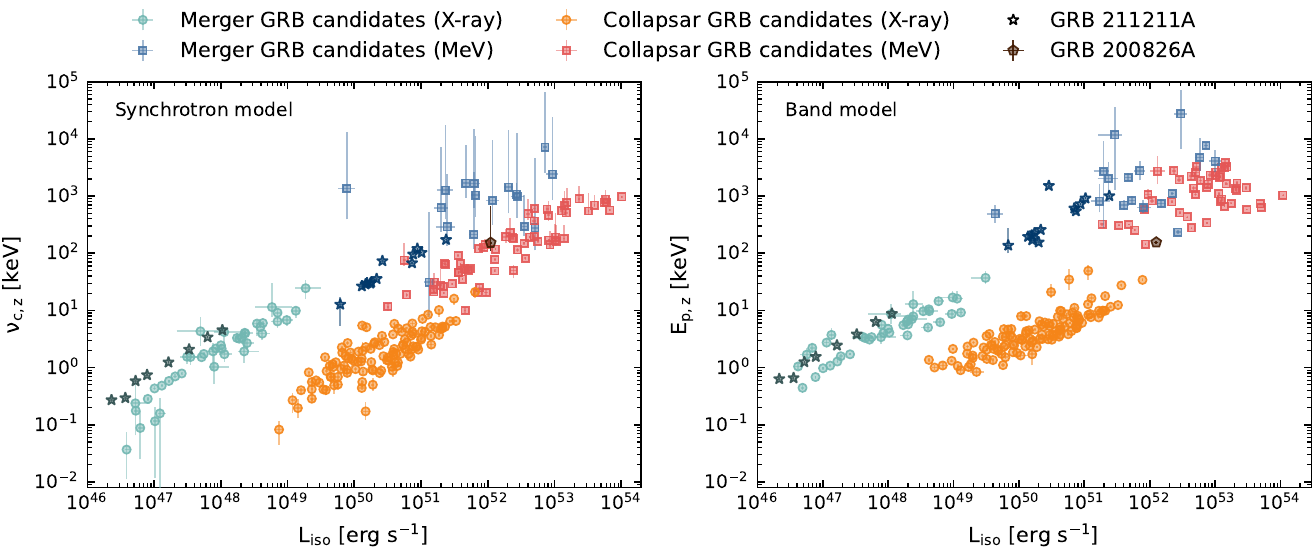}
    \caption{Spectral-luminosity correlations for merger-driven and collapsar-driven GRB candidates, derived by fitting the physical synchrotron model ($\nu_{c,z} - L_{\rm iso}$, left) and the empirical band function ($E_{p,z} - L_{\rm iso}$, right) to their time-resolved spectra. Circles represent parameters derived from XRT and BAT early X-ray observations (teal: merger-driven candidates; orange: collapsar-driven candidates), while squares represent parameters derived from MeV observations with GBM (blue: merger-driven candidates; red: collapsar-driven candidates). GRB~211211A is highlighted by star symbols, while GRB~200826A is marked in brown.}
   \label{fig:relations}
\end{figure*}

We performed a systematic analysis of the early X-ray emission of long GRBs detected by the \textit{Neil Gehrels Swift Observatory} \citep[hereafter \textit{Swift};][]{2004ApJ...611.1005G}. We focused on the steep decay phase, generally interpreted as the final stage of the prompt emission, which is characterized by an X-ray light curve segment steeper than $t^{-3}$ and shows a hard-to-soft spectral evolution \citep{2006ApJ...642..354Z}. We selected a sample of {23} long GRBs with measured redshift, $z$, that exhibit a clear steep decay phase in their X-ray light curve with a peak flux above $10^{-8}\,\mathrm{erg\,cm^{-2}\,s^{-1}}$ (Appendix~\ref{Appendix:catalog}). These events constitute our sample of collapsar-driven GRB candidates in the X-ray band.

For each GRB, we performed a time-resolved spectral analysis of data from the \textit{X-Ray Telescope} \citep[XRT; $0.3-10$~keV;][]{2005SSRv..120..165B} and the Burst Alert Telescope \citep[BAT; $15-150$~keV;][]{2005SSRv..120..143B} during the steep decay emission. We modeled the observed flux decay and spectral evolution as produced by the cooling of a non-thermal spectrum whose peak sweeps across the observational band towards lower energies. We first tested a physical synchrotron emission model \citep{2019A&A...628A..59O}, assuming the fast-cooling regime, in which the injection frequency, $\nu_m$, is higher than the cooling frequency, $\nu_c$. From this model, we derived the evolution of the bolometric flux, $F_{\rm bol}$, and of $\nu_c$, which corresponds to the spectral break. We then tested the empirical Band function \citep{1993ApJ...413..281B}, which provided the evolution of $F_{\rm bol}$ and of the spectral peak energy, $E_p$. In both cases, we accounted for neutral hydrogen absorption in both the Milky Way and the host galaxy. 
Further details of the data reduction and spectral analysis, as well as the results of the fits, are provided in Appendix~\ref{Appendix:X-ray_spectral_analysis}. 

The early X-ray sample of merger-driven GRB candidates is taken from our previous work \citep{2026A&A...708A.190I}. For the synchrotron model we adopted the spectral fits presented therein, whereas for the empirical model we refitted the spectra with the Band function, for consistency with the collapsar candidates.

To compare the X-ray steep decay properties with those of the MeV prompt emission, we studied GRBs detected by the \textit{Gamma-ray Burst Monitor} onboard \textit{Fermi} \citep[GBM; $\rm 8\,keV-40\,MeV$;][]{2009ApJ...702..791M}. The MeV data of the collapsar GRB candidates are taken from the multiple-bin sample of \cite{2025A&A...693A.156M}, who performed time-resolved spectral fits of the prompt emission of long GRBs using both the synchrotron and Band models. Their selection comprises the brightest long GRBs detected by GBM and is independent of the X-ray long GRB sample analyzed in this work.
The MeV data of merger-driven GRB candidates include GRB 211211A \citep{2025A&A...693A.156M} and a sample GBM-detected short GRBs with measured $z$. 
We fitted the $T_{90}$-integrated prompt emission spectra using the synchrotron model and the Band function (Appendix~\ref{Appendix:GBM_spectral_analysis}).

%%%%%%%%%%%%%%%%%%%%%%%%%%%%%%%%%%%%%%%%%%%%%%%%%%%%%%%%%%%%%%
\section{Spectral - luminosity correlations}
\label{sec:results}

Spectral - luminosity correlations during the MeV prompt emission are well established, both when spectra are fitted with an empirical model \citep[$E_{p,z}-L_{\rm iso}$;][]{2004ApJ...609..935Y} and with a physical synchrotron model \citep[$\nu_{c,z}-L_{\rm iso}$;][]{2025A&A...693A.156M}. Here, $E_{p,z}=E_p(1+z)$ and $\nu_{c,z}=\nu_c(1+z)$ are the rest-frame spectral peak energy and synchrotron cooling frequency, and $L_{\rm iso}=4\pi d_L^2F_{\rm bol}$ is the isotropic-equivalent luminosity, with $d_L$ indicating the luminosity distance of the source. Both relations have been extended to lower energies by analyzing the early X-ray emission of merger-driven GRB candidates \citep{2026A&A...708A.190I}. To test whether these correlations also hold during the X-ray steep decay of our collapsar GRB candidates, we derived the intrinsic properties from the best-fit spectral parameters. 

For the synchrotron model, we find that $\nu_{c,z}$ and $L_{\rm iso}$ are correlated, as shown in  the left panel of Fig.~\ref{fig:relations} (orange dots), and quantified by the Spearman rank correlation coefficient, {$\rho=0.80$}, with its associated chance probability, {$P_{\rm chance}=10^{-26}$}. In the $\nu_{c,z}-L_{\rm iso}$ plane, the population of collapsar-driven GRB candidates is clearly separated from the merger-driven GRB candidates analyzed in \cite{2026A&A...708A.190I} (teal points), which occupy a lower-luminosity region. Notably, this sample includes not only short GRBs and short GRBs with extended emission (EE), but also long GRBs associated with compact binary merger progenitors (e.g., GRB 211211A), which are correctly identified as merger-driven events. In the same panel, we show prompt emission data for both collapsar and merger-driven GRB candidates detected by \textit{Fermi}/GBM (red and blue squares, respectively). When fitted with the synchrotron model, the X-ray steep decay appears to be the natural extension of the MeV prompt emission for both GRB classes. Although the MeV spectra show some indication of a distinction between the two GRB classes, the X-ray spectra reveal a much stronger separation between the merger-driven and collapsar candidate samples.

For the Band model, the relation between $E_{p,z}$ and $L_{\rm iso}$ for the considered samples is shown in the right-hand panel of Fig.~\ref{fig:relations}. We find a strong correlation between these two quantities during the X-ray steep decay of collapsar GRB candidates, with {$\rho=0.86$ and $P_{\rm chance}=3\times 10^{-39}$}. Although the separation between merger- and collapsar-driven GRB candidates is no longer evident in the MeV prompt emission, the X-ray data still allow the two populations to be distinguished. Unlike the synchrotron model, the Band model does not reveal a continuity between the prompt emission and the X-ray steep decay of the collapsar GRB candidates.

%%%%%%%%%%%%%%%%%%%%%%%%%%%%%%%%%%%%%%%%%%%%%%%%%%%%%%%%%%%%%%
\section{Diagnostic}

\begin{figure}[t!]
   \centering
      \includegraphics[width=1\linewidth]{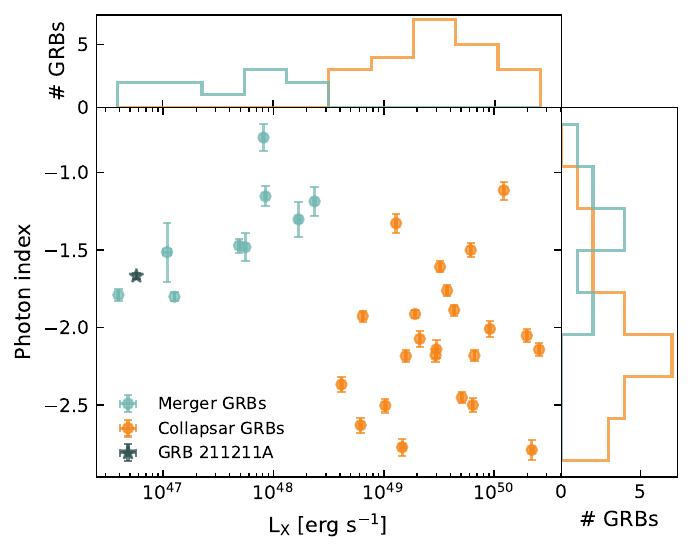}
    \caption{Photon index as a function of X-ray luminosity, derived from an absorbed power-law fit to the time-integrated spectrum of the steep-decay phase in the $0.5-4$ keV energy band. Teal and orange dots represent candidate merger-driven and collapsar-driven GRBs, respectively. GRB 211211A is highlighted with a star symbol. The right and top panels show the projected distributions of the photon index and X-ray luminosity.}
   \label{fig:Lx_time_integrated}
\end{figure}

The two relations identified using early X-ray and MeV observations (Fig.~\ref{fig:relations}) can provide a powerful tool for classifying GRB progenitors. Ideally, when the steep decay of a new burst is detected in X-rays, it could be placed in the $E_{p,z}$ - $L_{\rm iso}$ plane and classified accordingly. In practice, this is not always feasible, as such detailed time-resolved spectral analyses cannot be performed for all bursts. 
For example, the \textit{Wide-field X-ray Telescope} (WXT) onboard EP is sensitive only to a narrow energy range in the soft X-rays (0.5–4~keV), restricting spectral fitting to simple power-law models. This makes it difficult to reliably constrain the bolometric flux and spectral peak energy. Moreover, the lobster-eye optics employed by WXT are intrinsically less sensitive than the Wolter-I optics used by XRT, further limiting the quality of the spectral analysis.

To establish a realistic observational scenario for the current WXT, we reanalyzed the full sample of merger-driven and collapsar GRB candidates, restricting the analysis to the WXT energy band and developing a simpler observational diagnostic of the GRB progenitor.
We extracted the time-integrated XRT spectra during the steep decay phase and fitted them with an absorbed power-law model in the $0.5 - 4$ keV energy band, deriving the photon index and the intrinsic flux, $F_{\rm X}$, for each burst (Appendix \ref{Appendix:EP_sources}). We then plotted the photon index ($\Gamma$, defined with the convention $N_E \propto E^{\Gamma}$), which quantifies the spectral hardness, against the X-ray luminosity, $L_{\rm X}=4\pi d_L^2F_{\rm X}$.

Remarkably, the two populations remain well separated even with this simplified analysis, with collapsar-driven GRB candidates being more luminous and overall spectrally softer (Fig.~\ref{fig:Lx_time_integrated}). The same distinction is also found using time-resolved spectra (Fig.~\ref{fig:Lx_time_resolved}), showing that the physical spectral-luminosity relations observed during the steep decay can be approximated by $\Gamma-L_{\rm X}$ relations. This provides a practical diagnostic for classifying WXT fast X-ray transients (FXTs) exhibiting a steep decay phase, provided that they are associated with on-axis GRBs.
As discussed in Appendix~\ref{Appendix:EP_sources}, this approach correctly identifies EP 240801A \citep{2025ApJ...988L..34J}, a WXT-detected FXT associated with a GRB and a SN \citep{2026ApJ...996L..29V}, as a collapsar candidate.

%%%%%%%%%%%%%%%%%%%%%%%%%%%%%%%%%%%%%%%%%%%%%%%%%%%%%%%%%%%%%%
\section{Discussion and conclusions}
\label{sec:discussion}

We performed a time-resolved spectral analysis of the steep decay phase of collapsar GRB candidates, covering the soft-to-hard X-ray band. We found evidence for a spectral-luminosity correlation, independently of the adopted spectral model. The synchrotron model reveals a $\nu_{c,z}-L_{\rm iso}$ correlation, while the Band function shows a $E_{p,z}-L_{\rm iso}$ correlation (Fig.~\ref{fig:relations}). In both cases, the correlation of collapsar candidates is clearly separated from that of merger candidates analyzed in our previous work \citep{2026A&A...708A.190I}, highlighting their potential for identifying GRB progenitors.

Notably, in the X-ray band, long GRBs associated with compact binary mergers follow the spectral-luminosity relation of short GRBs and short GRBs with EE. This can help identify merger-driven GRBs, including events that would otherwise be classified as collapsar candidates based on their prompt-emission properties. For example, GRB 211211A exhibits a long-soft spectrum in the MeV band \citep{2022Natur.612..223R}, a characteristic typically associated with collapsar progenitors, while its steep-decay emission in the X-rays is consistent with the spectral-luminosity relation of merger-driven candidates.

The distinction between the two GRB classes during the X-ray steep-decay phase becomes even more striking when compared with the corresponding relations observed in the MeV prompt emission \citep{2004ApJ...609..935Y,2025A&A...693A.156M}, which show little or no evidence of separation. The synchrotron spectral-luminosity relations appear to extend continuously from the MeV prompt emission to the X-ray steep decay, suggesting a common origin for these two phases in both GRB populations. This continuity is less evident for the collapsar candidate sample when the spectra are fitted with the empirical Band function.

The simplified $\Gamma-L_{\rm X}$ diagnostic, developed by fitting the time-integrated spectra of the steep decay phase with an absorbed power-law model, makes our results applicable to current soft X-ray monitors, despite their limited energy coverage and sensitivity. In the $\Gamma-L_{\rm X}$ plane, merger-driven and collapsar GRB candidates remain clearly separated during the steep decay phase, even when the luminosities are evaluated in the $0.5 - 4$~keV energy band (Fig. \ref{fig:Lx_time_integrated}). Therefore, provided that the detected FXT is associated with on-axis GRB, this relation offers a practical method for classifying WXT sources.

The application of these relations requires the same selection criteria adopted for our analysis: (1) evidence that the newly-detected transient is a GRB (e.g., through the observation of hard X-ray or MeV prompt emission), and (2) presence of an X-ray steep decay phase with a peak flux $>10^{-8}\,\rm erg\,cm^{-2}\,s^{-1}$ in the 0.3–10 keV energy band. A measurement of the source redshift is also needed to place strong constraints on the progenitor.
Nevertheless, even when the redshift is unknown, the progenitor channel may still be constrained by evaluating the intrinsic spectral and luminosity properties over a range of possible redshifts and overplotting the resulting track in the relation plane.

Although the detection rate of such events inferred from the \textit{Swift} catalog may appear low, this might partly reflect the fact that XRT could observe the steep decay phase only after the spacecraft slewing, which typically requires $60-90$ s after the BAT trigger. In contrast, WXT can detect the steep decay emission from the earliest times, potentially providing a larger population of events for which these relations can be used. Moreover, the identified relations may remain applicable to the steep decay phase when these selection criteria are relaxed, or to prompt emission observations in the soft X-ray domain, which will be routinely provided by the WXT survey. 

Our results demonstrate that the early X-ray emission of GRBs provides a valuable and previously unexplored diagnostic of their progenitor channels. The separation of merger-driven and collapsar-driven GRB candidates in the \(\Gamma-L_{\rm X}\) plane during the steep decay phase establishes a promising observational criterion for distinguishing between these populations. With the growing sample of GRBs and FXTs detected by WXT, this approach will offer an opportunity to systematically investigate their progenitors.

\section*{Data availability}

The \textit{Fermi} and \textit{Swift} data are publicly available. Details on the GRB samples, results of the spectral analysis, and data points of the spectral-luminosity relations are available at the following link:
\url{https://zenodo.org/records/23613197}

\begin{acknowledgements}
The authors thank P. Jonker, G. Ghirlanda, and R. Cesarano for the fruitful discussions.
This work made use of data supplied by the UK Swift Science Data Centre at the University of Leicester. 
SR, MB, and BB acknowledge the ACME project, which has received funding from the European Union’s Horizon Europe Research and Innovation program under Grant Agreement No. 101131928. 
AM acknowledges support by funding from the Italian Space Agency, contract ASI/INAF n.I/004/11/6.
This research acknowledges support from the ASI-GSSI contract n. 2025-5-U.0: “Gamma-ray bursts: a probe of multi-messenger and extreme Astrophysics”. MER is supported by the ”la Caixa” Foundation Junior Leader Fellowship (ID
100010434). The fellowship code is LCF/BQ/PI25/12100030. This work was also
partly supported by the Spanish program Unidad de Excelencia Maria de Maeztu
CEX2020-001058-M, financed by MCIN/AEI/10.13039/501100011033, and by the
MaX-CSIC Excellence Award MaX4-SOMMA-ICE.

\end{acknowledgements}

%%%%%%%%%%%%%%%%%%%%%%%%%%%%%%%%%%%%%%%%%%%%%%%%%%%%%%%%%%%%%%

\bibliographystyle{aa}
\bibliography{bibliography}

\begin{appendix}

\section{X-ray sample selection}
\label{Appendix:catalog}

\begin{figure}[t!]
   \centering
   \includegraphics[width=\linewidth]{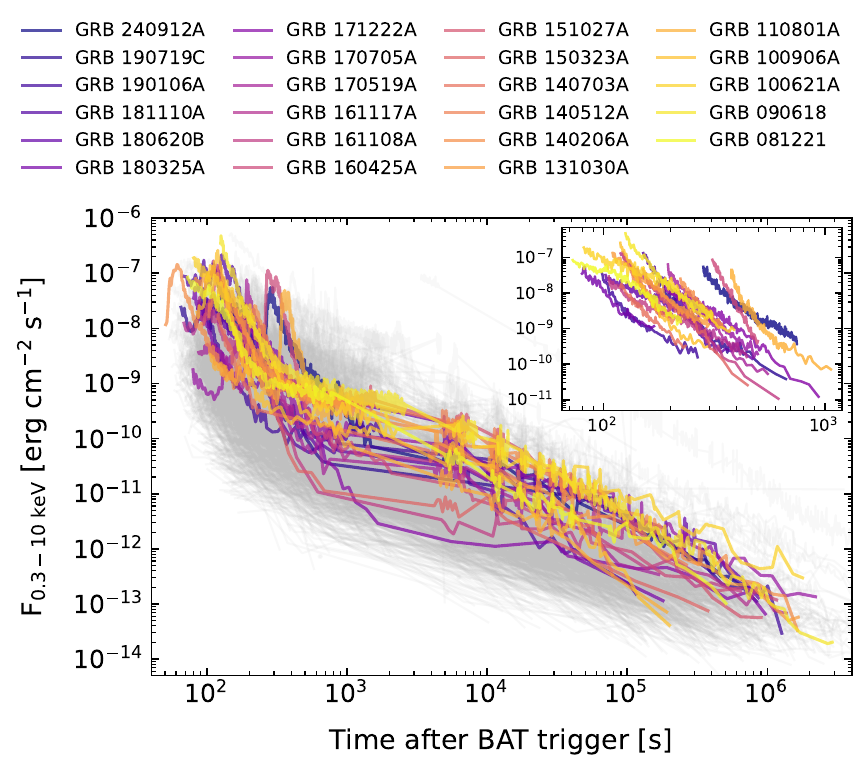}
   \caption{XRT flux light curves (0.3–10 keV) of the selected collapsar GRB candidates exhibiting a bright X-ray steep decay phase (colored curves), compared to the rest of the long GRB population (grey curves). The inset highlights the XRT steep decay phase identified for each selected GRB.}
   \label{fig:sample}
\end{figure}

To select our sample of collapsar GRB candidates in the X-ray band, we started from the \textit{Swift}/XRT ($0.3-10$~keV) catalog\footnote{\url{https://www.swift.ac.uk/xrt_live_cat/}}, which comprises 1718 GRBs detected up to the end of 2025. By cross-matching it with the \textit{Swift}/BAT ($15-150$~keV) catalog\footnote{\url{https://swift.gsfc.nasa.gov/results/batgrbcat/}}, we selected long-duration GRBs, defined as those with a prompt emission duration of $T_{90}^{\mathrm{BAT}} > 2~\mathrm{s}$. We excluded short GRBs with EE and long GRBs that likely originated from compact binary mergers (GRB 060614 and GRB 211211A). This initial selection yielded 1281 GRBs, of which 1079 were detected by XRT within $10^3$~s of the BAT trigger.
To ensure sufficient statistics for time-resolved spectral analysis, we restricted the sample to GRBs with a peak flux in the XRT light curve\footnote{\label{fn:XRT_lc}\url{https://www.swift.ac.uk/xrt_curves/}} \citep{2007A&A...469..379E} larger than $10^{-8}$~erg~$\rm s^{-1}~cm^{-2}$, yielding 176 GRBs. Since we aim to measure intrinsic properties, we then selected GRBs with measured spectroscopic redshifts (taken from the Greiner GRB catalog\footnote{\url{https://www.mpe.mpg.de/~jcg/grbgen.html}} or from the literature), resulting in a sample of 81 GRBs.
Finally, we required a clear steep decay phase in the XRT light curve ($\propto t^{-3}$ or steeper) with an initial flux $>10^{-8}$~erg~$\rm s^{-1}~cm^{-2}$. The steep decay has to occur within $T_{100}^{\mathrm{BAT}}$ to ensure that it is associated with the prompt emission rather than with late-time engine activity. We also required a hard-to-soft spectral evolution, identified through the evolution of the XRT hardness ratio, in order to minimize possible afterglow contamination. 
{If more than one steep decay phase satisfying the above criteria was identified in a single GRB, we selected the last one.} Our final sample consists of {23} collapsar GRB candidates with a bright X-ray steep decay phase, shown in Fig. \ref{fig:sample}, where they are compared with the full XRT-detected long GRB population.

Among the 81 GRBs retained after applying the flux and redshift
selection criteria, 23 satisfy all the additional requirements, necessary to obtain a clean sample for testing the presence of spectral-luminosity correlations during the X-ray steep decay phase.
Of the remaining 58 GRBs, 28 do not exhibit a steep decay
phase, while the other 30 show a steep decay that does not meet
our selection criteria, either because it is contaminated by
prolonged flaring activity or because it occurs after
$T_{100}^{\mathrm{BAT}}$.

\section{X-ray data analysis}
\label{Appendix:X-ray_spectral_analysis}

\subsection{Time bins}

To perform the time-resolved spectral analysis of the steep decay phase, we rebinned the XRT light curve of each burst using a binning algorithm similar to that of \cite{2021NatCo..12.4040R}, which we briefly summarize below.
Starting from the peak preceding the steep decay, consecutive bins of the XRT count-rate light curve\footref{fn:XRT_lc} were iteratively merged until (1) all count rates within the candidate bin were mutually consistent within $5\sigma$, (2) the accumulated counts exceeded a minimum threshold $N_{\rm min}$, and (3) the total bin duration was at least 5~s.
We chose $N_{\rm min}$ individually for each source according to its brightness. Typical values of $N_{\rm min}$ ranged between 2000 and 10000 in the Windowed Timing (WT) mode and around 1000 in the Photon Counting (PC) mode. This approach ensures finer time resolution during bright phases while maintaining sufficient counting statistics at lower flux levels. The BAT light curve was binned according to the previously defined time intervals for the XRT light curve.

\subsection{XRT and BAT data reduction}
\label{sec:X-ray_data_reduction}
For each defined time interval, we extracted the XRT source and background spectra in the corresponding WT or PC mode, together with the redistribution matrix files (RMFs) and ancillary response files (ARFs), using the automatic online tool provided by the \textit{Swift} Science Data Center for spectral analysis \citep{2009MNRAS.397.1177E}. We extracted the BAT spectra using the HEASOFT package (v6.33.1) following the standard procedure. The BAT event files were downloaded from the \textit{Swift} data archive and processed with the \texttt{batgrbproduct} pipeline. The BAT spectra were produced with the \texttt{batbinevt} task and applied systematic corrections using the \texttt{batupdatephakw} and \texttt{batphasyserr} tasks. The response matrices were generated with the \texttt{batdrmgen} task. We used the latest calibration files available in CALDB (release 2023-07-25).  

\subsection{Spectral analysis}
\label{sec:X-ray_spectral_analysis}

For each GRB, we performed a time-resolved spectral analysis of XRT and BAT data in the $0.3-150$~keV energy range, using XSPEC\footnote{\url{https://heasarc.gsfc.nasa.gov/xanadu/xspec/}} \citep[v12.14.0b;][]{1996ASPC..101...17A}. We adopted Cash statistics (\texttt{cash}) for XRT data and Gaussian statistics (\texttt{chi}) for BAT data. We estimated 1$\sigma$ parameter uncertainties using the \texttt{error} command.

\subsubsection{Spectral models}
\label{sec:spectral_models}

We fitted the spectra with two different models. For the physical synchrotron model, we adopted the XSPEC table model developed by \cite{2019A&A...628A..59O}. In this model, the emission is produced by a population of nonthermal electrons accelerated into a power-law energy distribution, \(dN_e/d\gamma \propto \gamma^{-p}\), above a minimum Lorentz factor \(\gamma_m\). As the electrons cool via synchrotron radiation down to the cooling Lorentz factor \(\gamma_c\), the resulting photon spectrum is obtained by integrating the single-electron synchrotron emissivity over the electron energy distribution. The model has four free parameters: the synchrotron cooling frequency \(\nu_c\), the electron power-law index \(p\), the ratio \(\gamma_m/\gamma_c\), and the spectrum normalization. For the empirical model, we adopted the Band function \citep{1993ApJ...413..281B}, widely used to fit GRB prompt emission spectra. In particular, we used the XSPEC \texttt{grbm} model, and derived $\alpha$, the spectral slope below the peak, $\beta$, the spectral slope above the peak, and the peak energy $E_p = (2+\alpha) E_c$.  

We modeled the X-ray absorption from neutral hydrogen in both the Milky Way and the GRB host galaxy using two independent absorption components. We adopted the XSPEC models \texttt{tbabs} and \texttt{ztbabs}, assuming the elemental abundances reported by \citep{2000ApJ...542..914W}. The Galactic hydrogen column density along the line of sight, $N_\mathrm{H}$, was fixed to the value derived by \cite{2013MNRAS.431..394W}. The host galaxy absorption component, $N_\mathrm{H}(z)$, evaluated at the GRB redshift, was instead left free during the spectral fitting. For each burst, this parameter was tied across all time-resolved spectra under the assumption that the intrinsic absorption remains constant throughout the steep decay phase.
We used the XSPEC \texttt{cflux} convolution model to compute the unabsorbed bolometric flux $F_{\rm bol}$ of each spectrum over the $10^{-3}-10^4$~keV energy range.

\subsubsection{Spectral fitting routine}

\begin{figure*}[t!]
   \centering
      \includegraphics[width=0.46\linewidth]{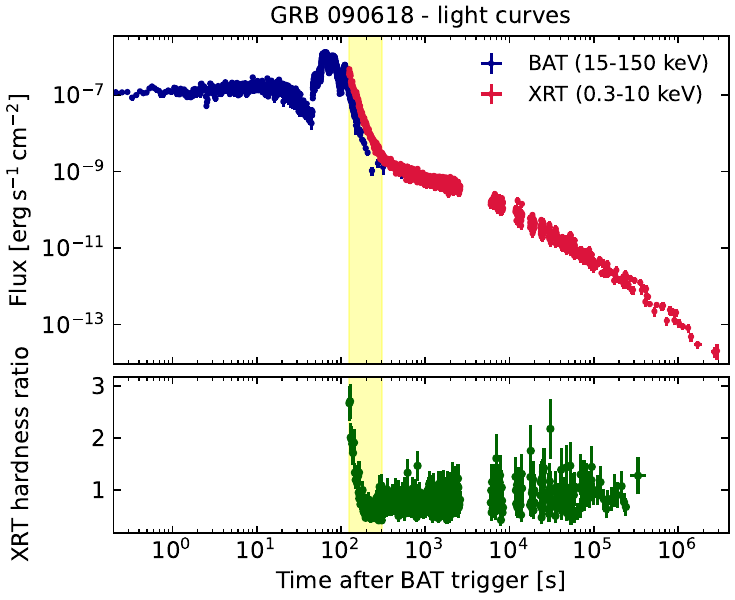}
      \hfill
      \includegraphics[width=0.505\linewidth]{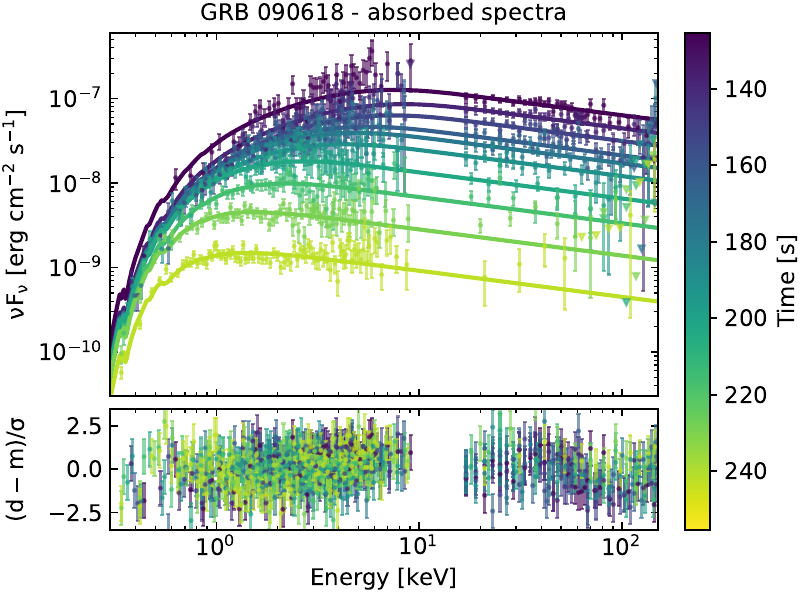}
    \caption{X-ray light curve (left panel) and time-resolved spectral evolution (right panel) of GRB~090618. The left panel shows the XRT and BAT observations selected to analyze the steep decay emission. In the upper panel, the blue and red data points represent, respectively, the BAT and XRT light curves. In the lower panel, the XRT hardness ratio is shown in green. The yellow shaded area highlights the time interval selected for the spectral analysis. The right panel displays the evolution of the absorbed spectrum of GRB~090618 in $ \nu F_\nu$ representation. The color scale indicates the different time bins. Data points from XRT and BAT are shown as dots, with triangles indicating upper limits. The solid curves represent our best-fit model. Residuals are displayed in the lower panel.}
   \label{fig:example}
\end{figure*}

\begin{figure*}[!htbp]
   \centering
   \includegraphics[width=0.95\linewidth]{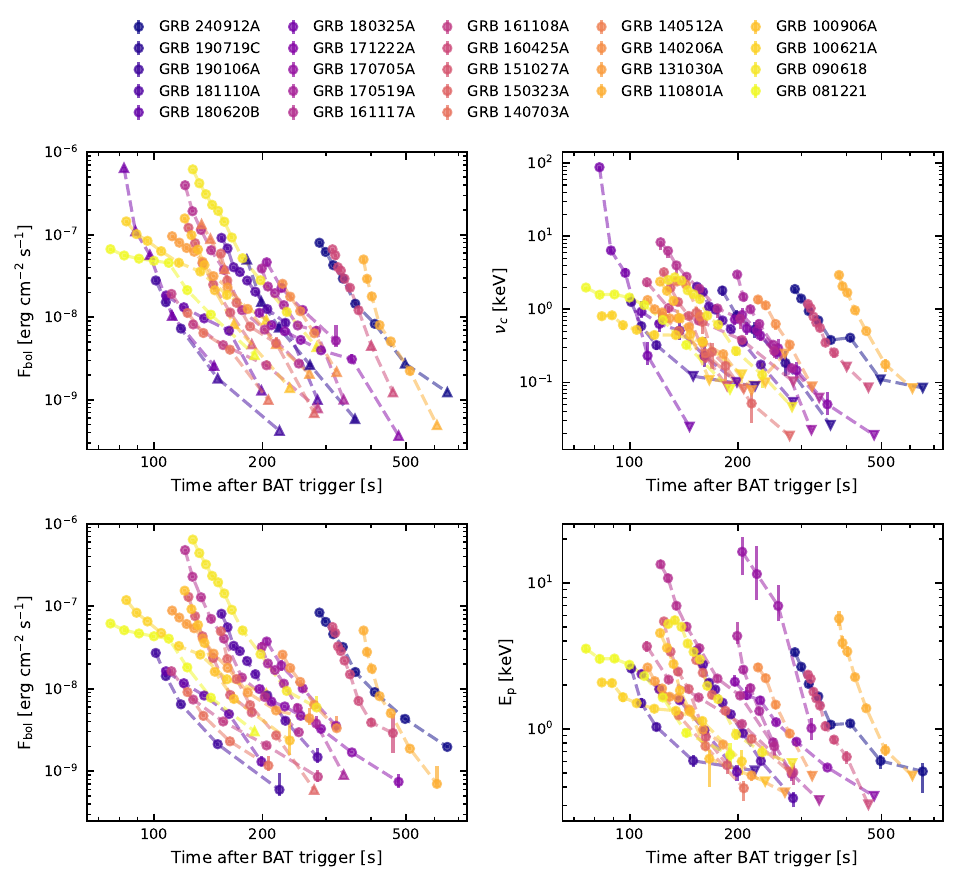}   

   \caption{Top: Evolution of the bolometric flux and of the synchrotron cooling frequency (top) for the X-ray steep decay emission of the long GRBs in our sample. Bottom: Corresponding evolution of the bolometric flux and peak energy for the Band model. Circles with error bars denote data points, while upward and downward triangles indicate upper and lower limits, respectively.}
   \label{fig:results}
\end{figure*}

We modeled the temporal and spectral evolution of the X-ray steep decay as the radiative cooling of a nonthermal spectrum whose peak progressively moves through the instrument energy bands. In our framework, the spectral shape remains unchanged during the steep decay phase, while the entire spectrum gradually shifts toward lower energies and decreases in flux. In Fig.~\ref{fig:example}, we show, as an example, the selected XRT and BAT observations \citep{2010A&A...519A.102E} of the steep decay emission of GRB~090618 (left panel), together with the comparison between the time-resolved spectra and the best-fit model (right panel). 

Since the XRT ($0.3$--$10$~keV) and BAT ($15$--$150$~keV) energy bands sample only a limited portion of the broadband spectrum at any given time, the spectral-shape parameters cannot generally be constrained independently in each time bin. We therefore performed a joint fit to all time-resolved spectra of each burst. The parameters defining the spectral shape ($p$ and $\gamma_m/\gamma_c$ for the synchrotron model, and $\alpha$ and $\beta$ for the Band model) were left free but tied across all spectra, whereas the characteristic energy ($\nu_c$ for the synchrotron model and $E_p$ for the Band model) and $F_{\rm bol}$ were allowed to vary independently in each time bin. A single value of $N_\mathrm{H}(z)$ was also fitted simultaneously to all spectra of the same burst.

The spectral parameters were allowed to vary in the following ranges: for the synchrotron model log($\gamma_m/\gamma_c$) $\in (0,2)$ (implying that all the shock-accelerated electrons efficiently cool via synchrotron losses, that is $\nu_m\geq\nu_c$); $p \in (2,5)$; and log($\nu_c${/keV}) $ \in(-4,3)$; for the Band model $E_p$ [keV] $\in$ (0.1, 200); $\alpha$ $\in (-1.9,0)$; and $\beta \in(-5,-2)$. For both models, the bolometric flux allowed range was log($\rm F_{bol}${/cgs}) $ \in (-15, -3)$.

Consequently, for a burst with $n$ time-resolved spectra, the joint fit contains $2n+3$ free parameters. For the synchrotron model, these consist of one value of $\nu_c$ and one value of $F_{\rm bol}$ for each spectrum, together with the common parameters $p$, $\gamma_m/\gamma_c$, and $N_\mathrm{H}(z)$. For the Band model, the free parameters are one value of $E_p$ and one value of $F_{\rm bol}$ for each spectrum, together with the common parameters $\alpha$, $\beta$, and $N_\mathrm{H}(z)$.
 
We applied this fitting procedure to all bursts in our sample, allowing us to track the temporal evolution of the bolometric flux and of the characteristic energy ($\nu_c$ for the synchrotron model and $E_p$ for the Band model). In some cases, we found that the last spectrum exhibited unexpected behaviour, such as a higher peak energy than that measured in the preceding time bin. This is likely due to contamination from an additional component (e.g., the afterglow). In these cases, we repeated the analysis excluding the last bin.

\subsection{Results}
The spectral analysis results for the early X-ray emission of the collapsar GRB candidates are shown in Fig. \ref{fig:results}.
For time bins in which the bolometric flux could not be reliably constrained (i.e., when its uncertainty exceeded the best-fit value), the measured 0.3–150 keV flux was adopted as a lower limit on $F_{\rm bol}$. This occurs predominantly toward the end of the steep decay phase, when the spectral peak shifts below the XRT energy band. 

For the synchrotron model, the bolometric flux drops by up to two orders of magnitude on a timescale of \(\sim 100\) s during the steep decay phase, while $\nu_c$ decreases by 1.5 orders of magnitude in the same time interval, as presented in Fig. \ref{fig:results}. For the Band model, a similar behaviour is observed for $F_{\rm bol}$ and $E_p$, with the latter spanning a higher energy range than $\nu_c$. 

For three bursts (GRB~190719C, GRB~180325A, and GRB~140512A), the evolution of a spectral break can be followed from the X-ray spectra, whereas the peak remains unconstrained: it lies above 150~keV in the first time bins, while the late-time spectra are too faint to constrain it. Consequently, in these cases, we can only derive lower limits on the peak energy and the bolometric flux. These data points are therefore not reported in Fig.~\ref{fig:relations}. In addition, since the Band is a peaked function, it cannot provide a reliable fit. We therefore report only the synchrotron fits for these GRBs in Fig.~\ref{fig:results}.

We note that, for GRB~180620B, fitting the time-resolved spectra assuming the cooling of a spectrum with a fixed spectral shape does not provide an adequate description of the spectral evolution. In particular, we find that the first bright spectrum (102--114~s) dominates the fit and consequently determines the spectral shape adopted for all subsequent spectra, introducing a bias in the inferred evolution. When this first spectrum is excluded, a different spectral shape and evolution are obtained, indicating a change in the spectral shape between the first and second time bins. Thus, a single fixed spectral shape cannot adequately describe the spectral evolution during this initial interval. We therefore perform the joint fits for this burst starting from 114~s onward.

\section{MeV data analysis}
\label{Appendix:GBM_spectral_analysis}
To compare the properties found for the X-ray steep decay phase to those of the MeV prompt emission, we compiled spectral analyses of \textit{Fermi}/GBM observations. The data for long GRBs, analyzed with both the synchrotron model and the Band function, were taken from \cite{2025A&A...693A.156M}. The short GRB spectra were fitted with the same models in this work. We selected from the GBM burst catalog all short GRBs detected by GBM with a measured redshift up to the end of 2025, and imposed that the 64 ms peak flux in the $50-300$ keV energy band has to be $>3.5 \rm \, cm^{-2}\,s^{-1}$, resulting in a sample of 18 bursts.

\subsection{GBM data reduction}
We downloaded the \textit{Fermi}/GBM data ($8\ {\rm keV} - 40\ {\rm MeV}$) from the Fermi GBM Burst catalog\footnote{\url{https://heasarc.gsfc.nasa.gov/W3Browse/fermi/fermigbrst.html}} of all the short GRBs in our samples, and performed a standard data reduction using the Fermi science tool \textsc{GTBURST}\footnote{\url{https://fermi.gsfc.nasa.gov/ssc/data/analysis/scitools/gtburst.html}}. In particular, we considered data from the two sodium iodide (NaI, $8-900$ keV) and one bismuth germanate oxide (BGO, $0.3-40$ MeV) detectors with best observational conditions (i.e., lowest separation angles).

The background analysis was performed through \textsc{GTBURST} by using, whenever the polynomial fit was converging, the background intervals reported by \textit{Fermi} collaboration \citep{vonKienlin2020}. In case this selection was leading to a non-convergent background polynomial fit, larger custom background intervals were selected.
We selected the source interval to coincide with the whole duration of the burst, measured through the $T_{90}^{\rm GBM}$ parameter reported in the Fermi GBM Burst catalog. The outputs of the data reduction are the source, background, and weighted response spectral files that are used for the spectral analysis through \textsc{XSPEC}.

\subsection{GBM spectral fitting routine}

For the fitting process, we used \textsc{pyXSPEC}\footnote{\url{https://heasarc.gsfc.nasa.gov/xanadu/xspec/python/html/index.html}}, the Python interface to the \textsc{XSPEC} spectral-fitting program. We implemented the Python-based software Bayesian X-ray analysis \citep[BXA,][]{Buchner2016}, which allows for Bayesian parameter estimation and model comparison through nested-sampling algorithms in \textsc{pyXSPEC}. 

We ignored the energy channels outside $8-900$ keV for the NaI detectors, as well as the $30-40$ keV band in order to avoid the iodine K-edge line at 33.17 keV \citep{Meegan2009}. We selected the energy range $\rm 300 \, keV - 40 \, MeV$ for BGO data.

For each dataset, we tested the same idealized physical synchrotron model described in Section~\ref{sec:spectral_models}.
We included the presence of a cross-calibration constant, \texttt{constant} in \textsc{XSPEC} notation, allowing for a $30\%$ variation for each dataset. We applied Poisson-Gaussian statistics (\texttt{pgstat}) to GBM data. For each model, we measured the logarithm of the $10 \, \rm keV - 10 \, \rm MeV$ bolometric flux in the source rest frame $\log F$ using the convolutional model \texttt{cflux} in \textsc{XSPEC}.

\subsection{Parameter estimation}

For each free parameter of the synchrotron model we defined broad and uninformative priors: $p \in (2, 5)$; $\log (F/{\rm cgs}) \in (-10, -3)$; $\log(\gamma_m/\gamma_c)\in (-1, 2)$ and $\log (\nu_c/\rm keV )\in (0,5)$. The analysis returns posteriors of the parameters fit to our dataset together with the Bayesian evidence. Similarly, for each free parameter of the Band model applied to the short GRBs sample, we define the following broad and uninformative priors: $\alpha \in (-6, 2)$, $\beta \in (-5, -1.9)$, $E_c\ [{\rm keV}] \in (8, 4 \times10^4)$ and $\log (F/{\rm cgs}) \in (-10, -3)$. 

Being interested in physical parameters too, we derived them from the inferred ones.  We defined the rest-frame synchrotron frequency associated with $\gamma_c$ as  $\nu_{c,z} = (1+z) \cdot 10^{\log \nu_c}$ and with $\gamma_m$ as $\nu_{m,z} = \nu_{c,z} \cdot (\gamma_m/\gamma_c)^2$. For the Band model, we computed the rest-frame peak energy,  $E_{p,z} = E_c \cdot (2+\alpha) \cdot (1+z)$. For each fitted model we also computed the flux, $F = 10^{\log F}$, and the total isotropic-equivalent luminosity, $L_{iso} = 4\pi d_L^2(z) \cdot F$. For each relevant parameter, we defined the best-fit value as the median on the posterior distribution, and lower and upper errors were derived from the 16th and 84th percentile of the posterior distribution, respectively.

\subsection{Results}

The intrinsic properties of the short GRB sample derived from the GBM spectral analysis are represented as blue squares in Fig.~\ref{fig:relations}. Since only one spectral break is typically observed in the GBM energy range for short GRBs \citep{2019A&A...625A..60R,2021A&A...652A.123T}, while the synchrotron spectrum is characterized by two breaks, $\nu_c$ is only weakly constrained, resulting in relatively large uncertainties. For the same reason, the cooling regime is also difficult to constrain; however, all bursts are consistent with the fast-cooling scenario.

\section{Einstein Probe sources classification}
\label{Appendix:EP_sources}

\begin{figure*}[t!]
   \centering
      \centering
      \includegraphics[width=0.759\linewidth]{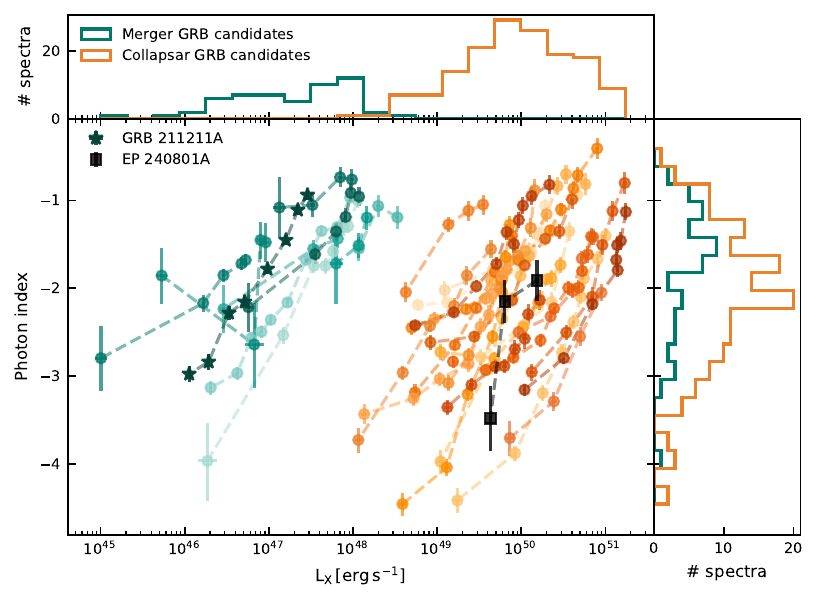}
      \caption{Photon index as a function of X-ray luminosity, derived from an absorbed power-law fit to the time-resolved spectra of the steep-decay phase in the 0.5–4 keV energy band. Teal and orange dots represent candidate merger-driven and collapsar-driven GRBs, respectively, with different shades denoting individual GRBs. GRB 211211A is highlighted by star symbols. The right and top panels show the projected distributions of the photon index and X-ray luminosity.  EP 240801A, used as a test case for classification, is marked with black squares.}
   \label{fig:Lx_time_resolved}
\end{figure*}

The two distinct spectral-luminosity correlations observed during the X-ray steep decay (Fig.~\ref{fig:relations}) can be used as a diagnostic for classifying GRB progenitors. However, deriving the evolution of $L_{\rm iso}$ and the spectral peak (or break) using the procedure described in Section \ref{sec:X-ray_spectral_analysis} requires broad energy coverage, spanning the soft-to-hard X-ray band throughout the steep decay phase. Such coverage is not available with EP/WXT, which detects transients in the relatively narrow 0.5–4 keV energy band. Within this range, the spectrum can generally be characterized only by a simple power-law model. To enable the direct classification of WXT-detected transients, we therefore characterize the spectral evolution during the steep decay phase using quantities derived from power-law fits.

If we consider the spectral evolution during the steep decay phase as described in Section \ref{sec:X-ray_spectral_analysis}, the spectrum is rigidly sweeping in the X-ray band towards lower peak energies and peak fluxes. In this scenario, when the X-ray spectrum in the WXT energy band is fitted with a power-law model, the sweep of the peaked spectral function appears as a transition from a hard to a soft photon index, accompanied by a decrease in X-ray flux. The photon index therefore gives an indication of the spectral peak energy, while the X-ray luminosity serves as a proxy for $L_{\rm iso}$.

To construct a more practical diagnostic for a direct comparison with the properties of WXT-detected sources (0.5--4 keV), we re-analyzed the steep decay emission of the collapsar GRB candidates selected in this work and the merger-driven candidates from \cite{2026A&A...708A.190I} using a simplified spectral fitting procedure. We adopted an absorbed power-law model to fit both the time-integrated and time-resolved spectra of the steep decay emission of each burst within the WXT energy band.

For the time-integrated analysis, we extracted the XRT spectrum of the steep decay phase (WT mode) following the procedure described in Section~\ref{sec:X-ray_data_reduction}. Each spectrum was then fitted in the 0.5--4 keV energy range with an absorbed power-law model (\texttt{tbabs*ztbabs*cflux*powerlaw} in XSPEC). The Galactic hydrogen column density along the burst line of sight, $N_{\rm H}$, was fixed \citep{2013MNRAS.431..394W}, while the intrinsic host-galaxy absorption at the burst redshift, $N_{\rm H}(z)$, was left as a free parameter. For each spectrum, we derived the photon index and the unabsorbed flux in the 0.5--4 keV band, which was used to compute the corresponding X-ray luminosity. The results of this analysis are presented in Fig.~\ref{fig:Lx_time_integrated} and show a clear separation between the collapsar and merger-driven candidate samples, providing quantities that can be directly compared with WXT detections.

To further investigate whether this separation is preserved in the spectral evolution under this simplified approach, we also fitted the power-law model to the time-resolved spectra. We adopted the same time binning and XRT spectra used for the analysis described in Section~\ref{sec:methods}. For each burst, we performed a joint fit of all time-resolved spectra using the same model as in the time-integrated analysis, with the host absorption parameter $N_{\rm H}(z)$ left free but common across all spectra. For each time-resolved spectrum, we derived the photon index and the unabsorbed 0.5--4 keV flux, and obtained a single $N_{\rm H}(z)$ value for each burst. The unabsorbed flux was then used to compute the X-ray luminosity. The results of this analysis are presented in Fig.~\ref{fig:Lx_time_resolved}, confirming the separation between the collapsar and merger-driven candidate samples. The $\Gamma-L_X$ relations in Fig.~\ref{fig:Lx_time_resolved} can be considered an approximation of the spectral-luminosity relations identified during the steep decay (Fig.~\ref{fig:relations}).

Given the lower sensitivity of WXT compared to XRT in the 0.5–4 keV band, we generally expect larger uncertainties in the derived spectral parameters. Nonetheless, the $\Gamma-L_X$ relations can still be used to classify WXT-detected FXTs, since the two GRB classes occupy well-separated regions of the parameter space, even when accounting for the larger uncertainties.

As a test case, we classified the EP 240801A, a FXT detected by WXT \citep{2025ApJ...988L..34J}. A coincident GRB was detected by \textit{Fermi}/GBM, and an associated SN was later identified \citep{2026ApJ...996L..29V}, providing strong evidence for a collapsar origin. Since the progenitor of this event is well established and it satisfies all the criteria required by our method, it provides an ideal test case for validating our classification scheme. We adopted the photon index and WXT flux density from the time-resolved spectral analysis of the X-ray steep decay ($103-198$~s) presented by \cite{2025ApJ...988L..34J}. We then computed the X-ray luminosity using $z=1.67$ \citep{2026ApJ...996L..29V}. The resulting time-resolved $\Gamma-L_{\rm X}$ measurements fall within the region of parameter space occupied by the collapsar GRB candidates (Fig. \ref{fig:Lx_time_resolved}), demonstrating that our method correctly classifies this event.

\end{appendix}

\end{document}